**The photovoltaic potential of tin perovskites revealed through layer-by-layer investigation of optoelectronic and charge transport properties.**

Mahmoud H. Aldamasy [1,2]‡, Artem Musiienko [1]*‡, Marin Rusu [1], Davide Regaldo [3], Shengnan Zho [1], Hannes Hampel [1], Chiara Frasca [1], Zafar Iqbal [1], Ahmed Bedir[1,3], Guixiang Li [1], Giuseppe Nasti [4], Meng Li [5], Jorge Pascual [6], Qiong Wang [1], Thomas Unold [1], Antonio Abate [1,4,7]*.

[1]Helmholtz-Zentrum Berlin für Materialien und Energie GmbH. Hahn-Meitner-Platz 1, 14109 Berlin, Germany.

[2] Egyptian Petroleum Research Institute, Nasr City, P.O. 11727, Cairo, Egypt.

[3]IPVF, Institut Photovoltaïque d'Ile-de-France, 18, Boulevard Thomas Gobert, 91120 Palaiseau, France.

[4]Department of Chemical, Materials and Production Engineering. University of Naples Federico II. Naples, pzz.le Vincenzo Tecchio 80, 80125, Naples Italy.

[5] Key Lab for Special Functional Materials of Ministry of Education, National and Local Joint Engineering Research Center for High-efficiency Display and Lighting Technology, School of Materials Science and Engineering, and Collaborative Innovation Center of Nano Functional Materials and Applications, Henan University, Kaifeng, 475004 China.

[6]Polymat, University of the Basque Country UPV/EHU, 20018 Donostia-San Sebastian, Spain.

[7]Department of Chemistry Bielefeld University, Universitätsstraße 25, 33615 Bielefeld, Germany

‡ Those authors contributed equally

* **corresponding authors:** antonio.abate@helmholtz-berlin.de, artem.musiienko@helmholtz-berlin.de

## Abstract

Tin perovskites are a promising environmentally friendly alternative to lead perovskites. However, the performance of tin perovskites is incomparable to lead perovskites for several reasons, including energy band mismatch with the charge transporting layers and high *p*-doping concentration due to tin oxidation. Researchers have found varying electrical and defect properties for tin perovskite devices, making it challenging to understand its operation limiting factors and unlock its full photovoltaic potential. In this work, we study the electrical, charge transport, and defect properties of the most prevailing $FASnI_3$-based photovoltaics devices. Using the kelvin probe method and photoelectron yield spectroscopy, we investigate the device stack layer-by-layer to construct the band diagram and quantify the interfacial energy band misalignment and each layer's contribution to the overall device performance. We found that the Bathocuproine buffer layer (BCP) has a major role in enhancing the open circuit voltage through blending with silver to form an intermediate energy level that dramatically enhances electron extraction. Using time-resolved surface photovoltage (trSPV), we study the interfacial charge transport dynamics at different charge extraction layers to understand the mechanism of charge extraction in tin perovskites. Based on extensive experimental findings, we developed a model of a solar cell that reveals the limiting factors of tin perovskite technology. For example, we identified a high recombination rate at the perovskite and hole transport layer (HTL) interface in the *p-p-n* architecture due to high excitation and low extraction. As a result, adapting the *n-p-p* architecture leads to higher extraction and less recombination. By creating a digital twin of the solar cell device, we formulate a roadmap for tin perovskite technology to achieve efficiencies exceeding 22%.

**Introduction**

Perovskite materials demonstrated outstanding performance in various photovoltaic applications due to their optimum charge transport properties and tunable bandgap. With a record efficiency of over 26%, [1] perovskite solar cells (PSCs) have matured to the point where it stands on the threshold of commercialization. [2,3] Nevertheless, these ambitions collide with several obstacles that may delay or, in the worst-case scenario, prohibit the market reach of PSCs. Among those are lead toxicity and bioavailability[4,5], which would be considered an imminent danger in case of cell rapture or cleavage. In this context, tin is the most suitable environmentally friendly lead substitute due to size and electronic similarity.[5] However, the best-performing Tin perovskite solar cells (TPSCs) show efficiencies over 10 percent lower than their lead counterparts and lag even further behind regarding stability and reproducibility.[6,7] One of the leading causes of the low efficiency is energy band misalignment, as shown in **Figure SI 1**.

Tin perovskites show a shallower valence band maximum (VBM) and conduction band minimum (CBM) than lead perovskites.[8] As a result, the energy bands of the most widely used charge transporting layers (CTLs) are misaligned with the VBM and CBM of the tin perovskite absorber material.[8,9] This energetic mismatch produces low charge extraction efficiency[10]. Consequently, tin perovskites show hampered photovoltaic performance compared to lead perovskites and their theoretical potential with an efficiency of up to 33%.[11,12] Practically, the tin perovskites community first adopted the regular *n-i-p* device architecture and then relied on metal oxides such as $TiO_2$ and $Nb_2O_5$ as electron transporting layers (ETLs) and spiro-OMeTAD as hole transporting layer (HTL). However, this device architecture suffered from chemical and physical complexities due to the reactivity of metal oxides. [13] With the emergence of fullerene application in PSCs [14], the tin perovskite solar cell device structure was switched to the inverted *p-i-n* architecture based on poly (3,4-ethylene dioxythiophene)-poly (styrene sulfonate) (with PEDOT: PSS) as HTL due to its easy processability and good wettability, and $C_{60}$/BCP as ETL. The later structure was globally adopted and matured to the point that it delivered a power conversion efficiency (PCE) of over 15% regardless of the considerable energy mismatch at both interfaces. [7,15] Therefore, it is crucial to manage the interfaces energetically to mitigate the interfacial losses.[16]

To transform the promising theoretical predictions of TPSCs into realistic devices, we need an in-depth electronic study of TPSCs. For example, we need an accurate energy band diagram to quantify the energy band offsets at each interface. Additionally, we have to understand and quantify the effect of the interfacial energy misalignment and *p*-doping concentration on the charge extraction dynamics and the device performance to set a bar for each parameter to achieve. Moreover, we need to compare the charge extraction dynamics and loss mechanisms of tin to lead perovskites to understand the electronic differences between both systems, which would help determine which strategy is exchangeable. Such detailed studies are required to fill the knowledge gap on the electronics and energetics of tin perovskites. However, to the best of our knowledge, there are no comprehensive studies on the electronic and charge transport properties of tin halide perovskites compared to lead perovskites. It is worth mentioning that measuring the precise energy level positions and interfacial charge transfer rates in TPSCs is challenging, especially on thin film stacks. The vacuum conditions and the energetic excitation beams could harm the sensitive tin perovskite films during measurements, emphasizing the need to explore suitable characterization techniques adapted to the unique nature of tin perovskites.

In this work, we use tin perovskite non-destructive techniques to study the photovoltaic, optoelectronic potential, and charge transport properties of tin perovskites through layer-by-layer investigation of the most prevailing tin perovskite device stack. We used an ambient-pressure Kelvin probe method (KP) and photoelectron yield spectroscopy (PYS) to construct the band diagram for tin perovskite devices under an inert $N_2$ atmosphere to quantify the energetic mismatch at each interface and to estimate the contribution of each layer to the overall $V_{oc}$ of the device. [16] Then, we accurately measured the p-doping concentration in the prepared FASnI3 films using the Hall effect, photoluminescence spectroscopy (PL), and KP-PYS. Furthermore, we used time-resolved surface photovoltage (trSPV) to study charge extraction rates and to form a comprehensive understanding of the interfacial charge extraction dynamics of different HTLs and ETLs including self-assembled monolayers (SAMs).[17-18]. Furthermore, we quantitatively analyze charge extraction efficiency and the mechanisms responsible for charge loss by conducting charge extraction and recombination simulations for tin and lead perovskites to understand the performance-limiting factors and the primary charge loss mechanisms in tin perovskites. Finally, we simulate the performance of tin perovskite devices as a function of *p*-doping concentration and CBO. Then, we identify the optimum value of each parameter to achieve the highest possible PCE. Our results introduce a comprehensive understanding of the energetic landscape in tin perovskites and provide a road map towards highly efficient tin perovskites with a PCE of over 20%.

## Results and Discussion:

### The doping concentration of $FASnI_3$

To demonstrate the limiting factors of TPSCs and construct a roadmap for further development, careful characterization is needed to determine doping density, free charge mobility, valence and conduction band energies of the perovskite and selective layers, bandgap energy, and bulk and surface recombination rates. Therefore, we applied a set of specialized experimental methods to characterize the layers and interfaces of the most common Tin perovskite device stack. The accurate determination of the *p*-doping level (hole concentration) in $FASnI_3$ is highly demanded to understand the electronic behaviour and charge carrier dynamics. However, the reported *p*-doping level of $FASnI_3$ is not accurately measured, and the values in the literature range from $10^{15}$ to $10^{18}$ $cm^{-3}$. [19,20] Therefore, we decided to measure the *p*-doping in neat $FASnI_3$ using the Hall effect, PL, and KP-PYS techniques. **Figures 1a and b** show the hole concentration of prepared $FASnI_3$ with 10% $SnF_2$ film. The Hall effect and PL techniques show the same concentration of $1.5 \times 10^{17}$ $cm^{-3}$. In assertive agreement with these values, on freshly prepared $FASnI_3$ thin films on quartz glass and indium tin oxide (ITO) substrates, the KP-PYS measurements reveal hole concentrations of $1.9 \times 10^{17}$ and $1.3 \times 10^{17}$ $cm^{-3}$, respectively. Those values are calculated with the determined $E_F$-$E_{VBM}$ energies of 0.01 and 0.08 eV from KP-PYS measurements (see next chapter) and hole effective mass in $FASnI_3$ of $0.05m_0$ as reported by Xie et al. The consistency of the measured hole concentration from Hall effect, PL, and KP-PYS demonstrates that the 0.05 $m_0$ hole effective mass, which fits the experimental data from these different measurement methods, is the most accurate value compared to the rest reported in the literature.[21]

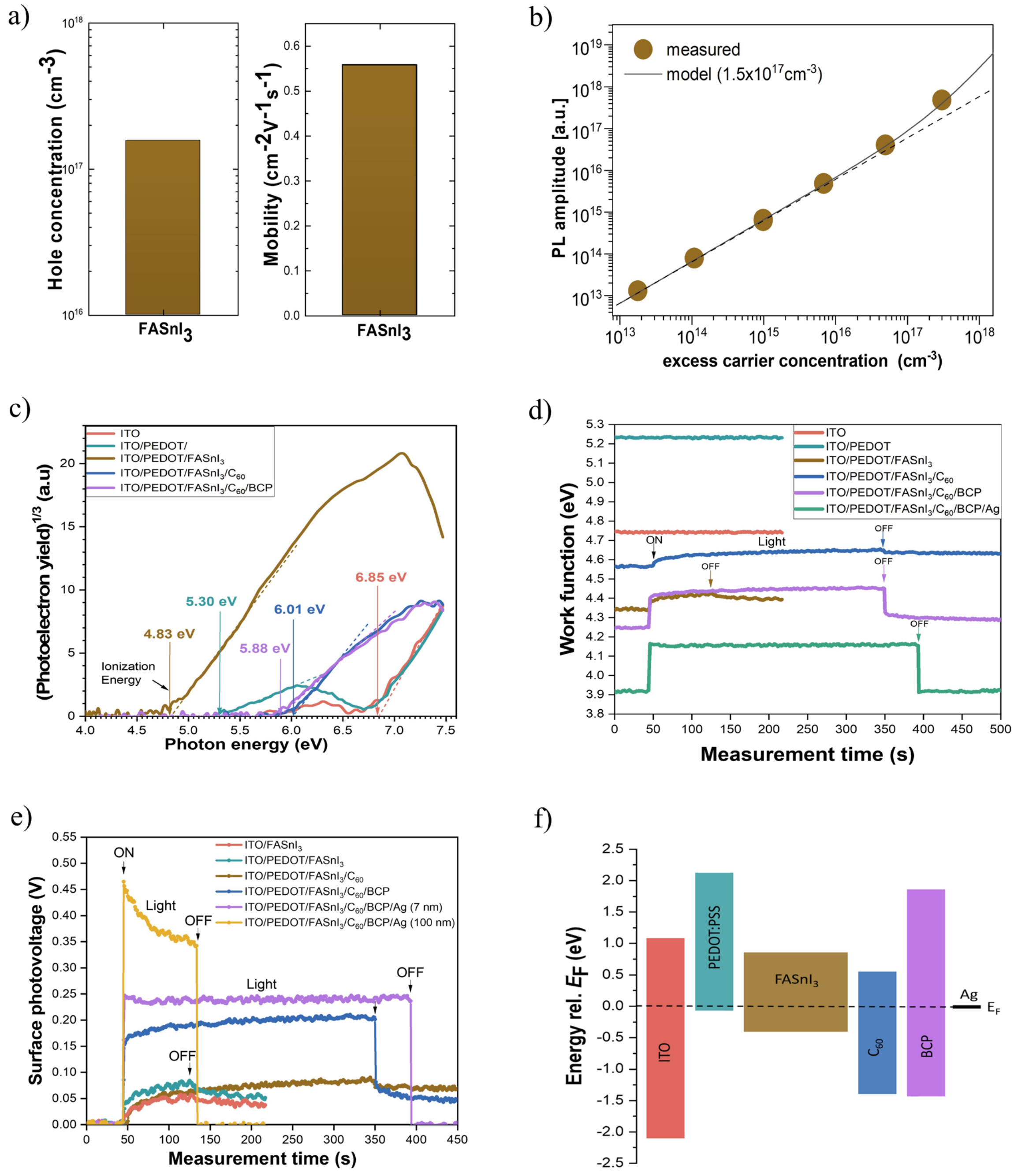

**Figure 1: The electronic and energetic landscape of the ITO/PEDOT:PSS/$FASnI_3$/$C_{60}$/BCP/Ag solar cell**. (a) Charge carrier concentration and mobility of $FASnI_3$ thin films. (b) Doping concentration was determined by fitting the PL amplitude as a function of excess carrier concentration from TRPL measurements. (c) Photoelectron yield spectra of each layer of the ITO/PEDOT:PSS/$FASnI_3$/$C_{60}$/BCP/Ag device stack with the revealed ionization energies from layer-by-layer studies. (d) Time-dependent WF measurements in the dark and under illumination measured layer-by-layer from ITO to the complete solar cell stack. (e) SPV evolution from the ITO film to the full device stack ITO/PEDOT:PSS/$FASnI_3$/$C_{60}$/BCP/Ag revealed by KP layer-by-layer measurements. (f) The energy band diagram of the ITO/PEDOT:PSS/$FASnI_3$/$C_{60}$/BCP/Ag structure with respect to Fermi level ($E_F$) is plotted based on the data determined from KP and PYS measurements.

**Energy band diagram of the ITO/PEDOT:PSS/$FASnI_3$/$C_{60}$/BCP/Ag solar cell**

To understand the efficiency losses related to CTLs in TPSCs, we analyzed the interfacial energy bands alignment between the absorber layer and different CTLs in addition to the charge carrier extraction dynamics. The determined positions of the conduction and valence bands of different tin perovskite compositions and CTLs scatter significantly in the literature. This is partly explained by the difficulty of getting consistent measurements due to challenges like $Sn^{2+}$ sensitivity to oxygen, beam damage during high-energy photoelectron analysis, and composition changes under high vacuum conditions. Therefore, We have performed the KP-PYS measurements under an inert N2 atmosphere under normal conditions and employed low energy photons in the 3.4 – 7.5 eV range. Based on the aforementioned measurements, we built the band diagram for the studied ITO/PEDOT:PSS/FASnI3/C60/BCP/Ag solar cell device. The Fermi level position is found from work function (WF) measurements by KP (**Figure 1d**), and the band bending and the contribution of each layer to the solar cell open-circuit voltage ($V_{OC}$) are found from the magnitude of the SPV signal derived from KP data in the dark and under illumination (**Figure 1e**). The VBM position is determined as the ionization energies found from PYS spectra, which were found by linear regression and extrapolation of the $Y^{1/3}(h\nu)$ curves in **Figures 1c and SI 2-5**, except that of Ag, which was determined separately from $Y^{1/2}(h\nu)$ plots valid for measurements on metals. Knowing the VBM position, the CBM position is found by adding the bandgaps of the respective thin films to the $E_{VBM}$. Assuming flat band conditions under the used 100 mW/cm$^2$ illumination, the doping concentration of the films is calculated. It is worth mentioning that the PYS measurements for the determination of the ionization energy ($E_i$) and the KP measurements for the determination of the WF, which determine the Fermi level ($E_F$) position with respect to the vacuum level, were performed with the same Kelvin tip on the same sample position. Note that the WF of the KP tip was checked each time before starting a new measurement. Furthermore, repeatedly performed PYS measurements have delivered the same ionization energy values denoting the chemical stability of the investigated thin film surface.[8,16,22-23]

**Figure 1d** shows the layer-by-layer WF measurements in the dark and under illumination of the TPSC device. The measurements were stable and reproducible over time, proving the prepared films' surface chemical and electronic stability. The light-induced WF variations in **Figure 1d** and the corresponding SPV development in **Figure 1e** were first observed after the deposition of the $FASnI_3$ film. Since no SPV was recorded from ITO and PEDOT:PSS films, the observed SPV amplitude of about 70 mV is attributed to the potential barrier at the $FASnI_3$ film surface due to band bending. The $C_{60}$ layer, although deposited as an ETL on top of the $FASnI_3$ absorber film, did not enhance the SPV amplitude. Note that the SPV measured from a single film would be attributed to the surface band bending, while an SPV measured from a *p-n* junction shall coincide with the solar cell device's $V_{OC}$ of the solar cell. With well-passivated interfaces and appropriate contacts, the $V_{OC}$ of a photovoltaic device is expected to approach the internal voltage (related to quasi-Fermi level separation) within the absorber, as indicated in **Figure 1e** (yellow curve), in which the SPV signal of the full device stack matches the $V_{OC}$ value obtained by the current density-voltage (*J-V*) curve [24,25] **(Figure SI 6)**. Therefore, we conclude that the $C_{60}$ layer has no significant contribution to the built-in voltage within the $FASnI_3$ absorber. The SPV of the device stack increases significantly only after the deposition of the next BCP film. A considerable increase in the SPV signal is observed after depositing a semi-transparent 7nm-thick Ag layer. Since the sample is illuminated through the Ag film, the intensity of the incident light is significantly decreased. Thus, the recorded SPV deviates from the potentially maximum achievable SPV signal. Nevertheless, these results show that the internal voltage within the $FASnI_3$ absorber increases essentially after the BCP deposition while reaching its maximum after the completion of the device structure with the Ag film. In **Figure 1e**, we observe a sharp increase of the SPV signal to its maximum value by switching on the light and a sharp decrease of the SPV after switching off the illumination only after completing the solar cell stack with the Ag film. This shows optimum charge separation occurs at the back contact after the Ag film deposition. Although the SPV curve for the BCP completed device demonstrates an improved transfer of electrons from the absorber film to the BCP surface, the long SPV relaxation time after the illumination is switched off denotes the presence of surface trap states.

With the data obtained from KP and PYS measurements, we plotted the complete band diagram of the ITO/PEDOT:PSS/$FASnI_3$/$C_{60}$/BCP/Ag device stack under study in **Figure 1f**. We aligned it to the Fermi level for discussion of the current transport and to the vacuum level, as shown in **Figure SI 7,** to collect the data on electron affinity (EA), Fermi level position, and ionization energy of all the layers in the device stack. The Fermi level-aligned band diagram reveals the offsets of the energy levels at the perovskite HTL/CBM and the perovskite VBM/ETL interfaces. Those band offsets influence both the charge carrier extraction and recombination. For the PEDOT:PSS/$FASnI_3$ interface, we calculated a positive band offset $\Delta E_v = 0.42$ eV by using $\Delta E_v = (E_F^{pero} - E_i^{pero}) - (E_F^{HTL} - E_i^{HTL})$. Thus, under the assumption of a low density of recombination states at this interface, an effective extraction of holes from the $FASnI_3$ absorber would be expected. However, the extraction of holes into the degenerated ITO film bulk will be hindered by its superficial layer, which looks intrinsic, as depicted in **Figure 1f**. At the $FASnI_3$/$C_{60}$ interface, we calculate a negative band offset $\Delta E_c = -0.30$ eV by using $\Delta E_c = (EA^{ETL} - E_F^{ETL}) - (EA^{pero} - E_F^{pero})$. Consequently, no barriers exist for an effective electron extraction at this interface. However, since the SPV signal differed slightly from that of the $FASnI_3$ film and increased slowly under illumination while showing a long decay tail after the light switch-off, we assume a high density of trap states at this interface. The band structure of the BCP, as aligned to the $C_{60}$ energy levels, suggests a large barrier for the extraction of electrons. Considering the essential increase of the SPV signal with the addition of the BCP layer, one can assume the electronic transport through the BCP film is assisted by gap states. However, our KP-PYS investigations show that the BCP film's chemistry and electronics are strongly influenced by the deposition of the Ag layer, which we will discuss further in the next chapter.

## The electronic structure of the BCP/Ag interface

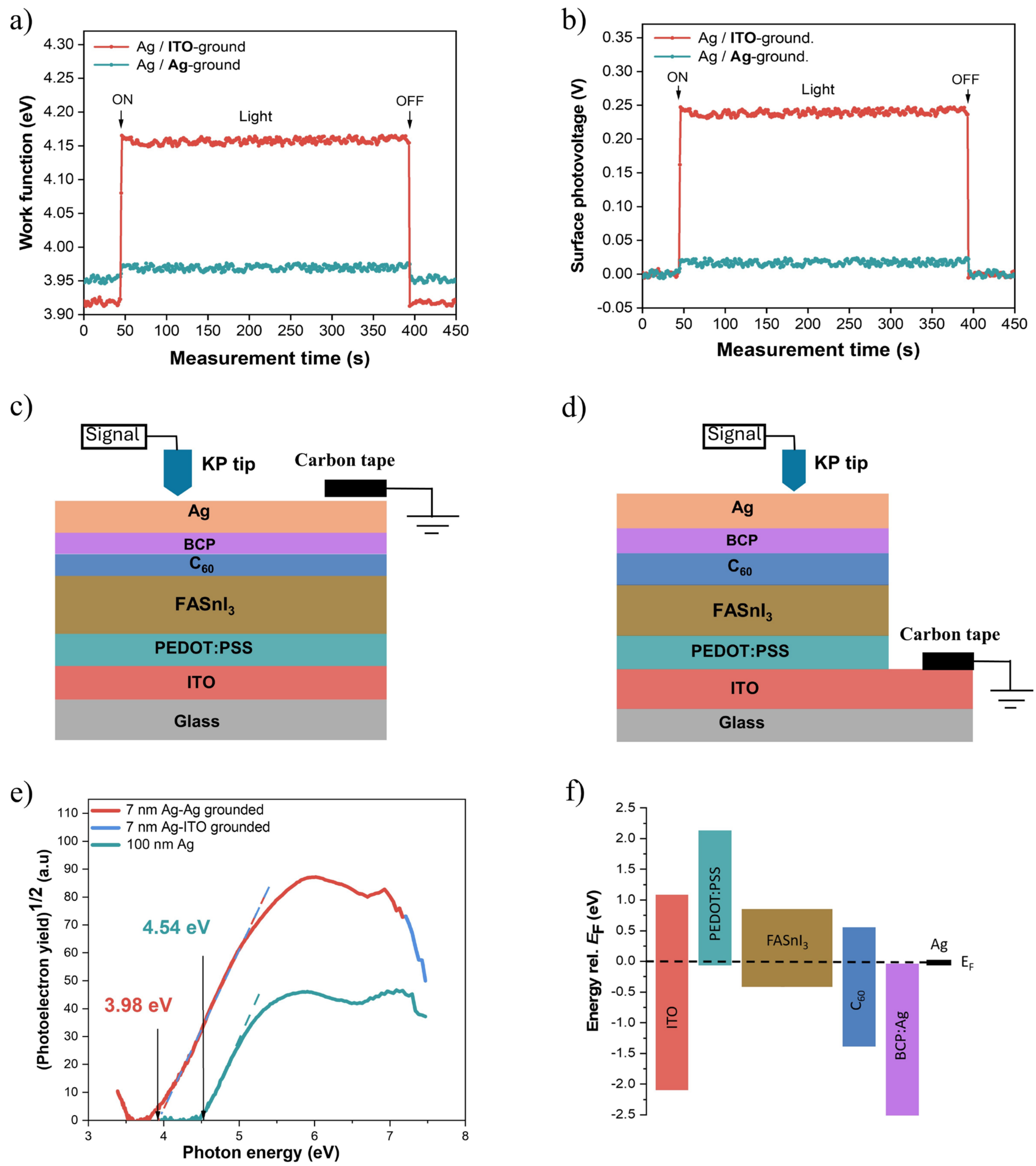

**Figure 2: Electronics of the Ag/BCP interface**. (a) WF and (b) surface photovoltage of the ITO/PEDOT:PSS/$FASnI_3$/$C_{60}$/BCP/Ag device from KP measurements by selectively grounding either the ITO or Ag films. (c) Schematic diagram of the KP-PYS Ag-grounded sample. (d) Schematic diagram of the KP-PYS ITO-grounded sample. (e) Photoelectron yield spectra of a 7-nm-thick Ag film on top of the BCP layer were measured by grounding the device stack either on the ITO or Ag films. (f) Proposed energy band diagram with a BCP:Ag blend layer with a LUMO level of 4.0 eV.

The BCP layer was inserted as a buffer layer between the ETL and the cathode to give rise to a cascade of energy levels for improved extraction of electrons and to create simultaneously a wide hole injection barrier that reflects holes, thus preventing the interface recombination.[26] However, the remarkable enhancement of the device SPV caused by the BCP/Ag stack motivated us to investigate the BCP/Ag interface to understand the origin of such an enhancement from the electronic point of view. Therefore, we prepared an entire device stack with a top Ag film of only 7 nm thickness. We carried out the KP measurements in the dark and under illumination by grounding the ITO film in the first measurement, while in the second measurement, we grounded the Ag film, as shown in **Figures 2 a-d**. The KP-PYS measurements were performed on the same sample position in both cases. Note that in the case of grounded ITO, the signals were recorded relative to the $E_F$ of the ITO thin film, while in the case of grounded Ag, the signals were recorded with respect to the $E_F$ of the Ag layer. From the dark KP measurements (**Figure 2a)**, we observe almost similar WF values between 3.92 and 3.95 eV for both types of sample grounding. Meanwhile, from the PYS measurement (**Figure 2e)**, we identify an $E_i$ value of 3.98 eV in both cases. These results show that the WF of an investigated thin film on a solar cell stack is properly determined by KP measurements, independent of whether the measurement is conducted by grounding the solar cell's ITO (bottom) contact or directly on the measured top layer. The SPV of the Ag-grounded measurement (**Figure 2b)** is attributed to the (minor) surface band bending of the top layer. In contrast, the SPV of the ITO-grounded plot is attributed to the $V_{OC}$ of the stack, which is related to the Fermi level splitting due to the generation of charge carriers under illumination in the $FASnI_3$ absorber thin film. Thus, both KP (dark) and PYS measurements reveal similar WF and $E_i$ values, respectively, as expected for a measurement on a metal or a degenerated semiconductor. However, the measured value of about 4.0 eV is (i) 0.56 eV lower than that of the 100 nm thick Ag reference film (**Figure 2e)**, (ii) 0.25 eV lower than the $E_F$ of BCP, and (iii) 1.88 eV lower than the $E_{i\text{-}BCP}$. Therefore, we suggest that the early deposited Ag diffuses into the BCP layer, forming a BCP:Ag blend with a WF of around 4 eV, which is in an optimum position for effective electron extraction from $C_{60}$.[27-28]

Lately, the concept of Ag:BCP blend layer was introduced by Ying et al. in tandem [29] and single junction [30] devices as a mediator between $C_{60}$ and indium-zinc oxide (IZO) as a transparent electrode. The target was to protect the $C_{60}$ layer from sputtering damage during the deposition of the top conductive electrode. They proved the formation of a charge transfer Ag:BCP complex through coordinate covalent bonds between Ag and N atoms of BCP during the evaporation process. Similar interactions between BCP and electrodes with a different metal, such as Mg, were also reported [31,32]. Ying et al. suggested that Ag:BCP complex forms gap states that work as channels to facilitate electron transportation to the electrode. [30]

On the other hand, our BCP/Ag interface investigation reveals similar energies of about 4.0 eV for the WF and ionization energy. Such findings indicate the formation of a degenerated BCP:Ag surface layer. Therefore, we propose that BCP:Ag is a complex with a Fermi level coinciding with its CBM position at 4.0 eV. This low WF contact layer provides a downward bending of the $C_{60}$ bands, thus enabling a smooth transfer of electrons from $C_{60}$ to the back Ag electrode, as shown in **Figure SI 8**. Conclusively, we suggest a modified energy band diagram as indicated in **Figure 2f**, in which a BCP:Ag blend with a WF ≈ 4 eV replaces the traditional BCP layer.

## Charge extraction dynamics at charge selective contacts:

### (a) Charges dynamics at the HTL buried interfaces.

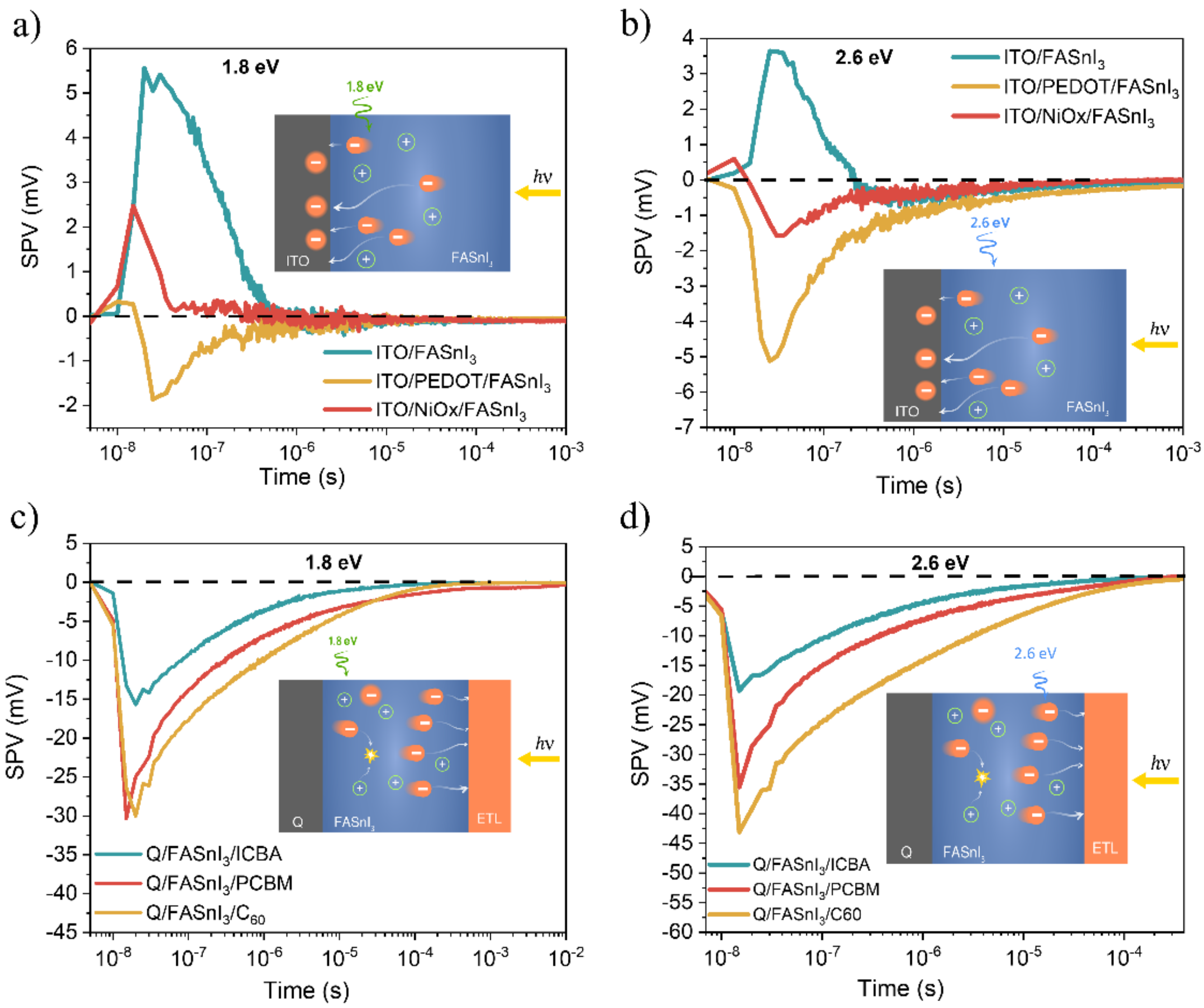

**Figure 3: Free electron and hole extraction dynamics in different ETL and HTL interfaces of TPSCs.** a& c) excitation using 688 nm laser pulse (1.8 eV), penetrating around 210 nm in a 250 nm-thick film and exciting electrons near the buried interface The films were deposied on ITO and quartz (Q) respectively.. b & d) Excitation using a 486 nm laser pulse (2.6 eV) penetrating up to 90 nm and exciting electrons in the bulk of the perovskite film.

After illustrating the energetic landscape in the most common TPSC stack, we investigated the ability of different CTLs to extract charges from the absorber layer by measuring trSPV. We illuminated the interfaces under study from the top side, as shown in **Figure 3**, and recorded charge dynamics from 5 ns to 10 ms. **Figures 3a and b** insets show the trSPV diagram of neat $FASnI_3$ at two different excitation energies, 1.8 and 2.6 eV, corresponding to penetration depths of 210 and 90 nm, respectively[33], as indicated in **Figure SI 9**. The sign of the SPV signal reflects the electron or hole separation in space. A positive (negative) sign of transient reflects the electron (holes) separation for the bulk of the thin film to its surface or interface, leaving the perovskite film positively (negatively) charged.[34] The amplitude and rise rate of the signal indicate the number of separated charges and thus express the efficacy of the charge extraction. The subsequent signal decrease indicates the recombination losses of the separated charge carriers. **Figure SI 10**.

We start by measuring the ability of ITO to extract charges directly from the $FASnI_3$ layer without any CTL. We found that the ITO/$FASnI_3$ interface shows a positive trSPV signal, which indicates an electron accumulation near/in ITO, as shown in **Figures 3a and b**. The peak signal decreased when we generated carriers further away from the interface with 2.6 eV photons (**Figure 3b**). This electron accumulation can be attributed to the electron trapping or transfer in ITO, as similar effects were observed in ITO/lead perovskite interface previously.[34] Further, we explored hole extraction dynamics in $FASnI_3$/HTLs interfaces. In contrast to ITO/$FASnI_3$, PEDOT:PSS/$FASnI_3$ interface shows negative signals across the perovskite film, revealing holes extraction in PEDOT:PSS. However, the signal is stronger when we generate carriers deeper in the bulk of the perovskite film (**Figure 3b**) than closer to the interface (**Figure 3a**). This change in signal amplitude demonstrates a high interfacial recombination rate or barrier at the PEDOT:PSS/$FASnI_3$ interface. This draws our attention to the significance of the interfacial passivation and alignment of the buried interface in TPSCs.[35] [36] Notably, PEDOT:PSS shows a small positive signal at the beginning, an early electron trapping across the interface before 10 ns, which is shortly overflooded by hole extraction (**Figure 3a**). This indicates that the process of charge extraction at the PEDOT:PSS/$FASnI_3$ interface is derived by a competition between two processes, hole extraction, and electron trapping, which could be attributed to energy bands misalignment and low charge selectivity of PEDOT:PSS. Nevertheless, the hypothesis on electron trapping is in perfect agreement with data measured at 2.6 eV for the PEDOT:PSS interface (**Figure 3b**). We did not observe any positive signal due to electron-hole generation and recombination far from the PEDOT:PSS interface and prevailed majority of free hole drift in the bulk of the *p*-type material.[37]

Next, we studied the hole extraction capabilities of NiOx and two carbazole-based SAMs (2PACz and MeO-2PACz, which stand for SAM3 and SAM2, respectively), which previously showed advanced hole extraction capabilities in lead-based perovskite cells[34] and were also tried for tin perovskites[38] The trSPV transient of NiOx shows a much less intense negative signal and much worse hole extraction capabilities than PEDOT:PSS. When free carriers are generated closer to the interface (with 1.8 eV photon), electron accumulation at the NiOx/$FASnI_3$ interface dominates, giving rise to a positive trSPV signal, as indicated in **Figure 3a**. At the same time, hole extraction becomes apparent only at 2.6 eV illumination, which is still affected significantly by the positive signal for electron accumulation. The detailed qualitative interpretation of charge transport for the NiOx interface is given in **Figure SI 11**. Conversely, only 2PACz showed signs of hole extraction at 2.6 eV illumination among SAMs, as shown in **Figure SI 12**. Similar to $NiO_x$, SAMs showed electron accumulation and low extraction capabilities compared to PEDOT:PSS. To validate the alignment effect on the hole extraction by SAMs, we added 20% $SnBr_2$ to deepen the valence band of tin perovskite.[8] Interestingly, we observe the improvement of hole extraction by $NiO_x$, 2PACz, and MeO-2PACz, which proves such HTLs could work with tin perovskites if the energy bands are appropriately aligned. In conclusion, these HTLs can be optimized for tin-perovskites with a broader bandgap for lead-free tandem applications.

### (b) Electron extraction dynamics at TPSCs

To explore electron extraction dynamics at perovskite/ETLs interfaces, we measured the trSPV of commonly employed fullerene derivatives indene-$C_{60}$ bisadduct (ICBA) and [6,6]-phenyl-$C_{61}$-butyric acid methyl ester(PCBM), and $C_{60}$ deposited on top of $FASnI_3$ as depicted in **Figures 3c and d**. In this structure, we illuminate from the top side. Hence, the higher excitation energy (2.6 eV) generates charge carriers close to the interface, while the lower energy laser (1.8 eV) excites near the bulk of the film, as shown in the inset of **Figure 3c and d**. We observe negative signals for all ETLs, which indicate an active electron extraction at the interface. The rapid increase in the signal amplitude suggests fast separation dynamics of electrons towards the ETL. Generally, $C_{60}$ shows a better extraction, especially when we excite near the interface. Interestingly, the charge extraction rate of the three ETLs follows the same order of charge mobility, which is 1.6, $6.1 \times 10^{-2}$, and $16.9 \times 10^{-3}$ $cm^2\ V\ s^{-1}$ for $C_{60}$, PCBM, and ICBA, respectively, and not the order of the conduction band offset (CBO) values. This correlation can explain the superior performance of the $C_{60}$-based devices regardless of the unmatched energy bands compared to ICBA and PCBM, keeping in mind also the vital role of BCP when coupled with $C_{60}$ to form one ETL as explained in a previous section.[39] In addition, the $C_{60}$ interface (as well as other ETLs) shows a much larger peak amplitude of the signal (-47 mV) than the PEDOT:PSS interface (-5 mV), which reveals much better electron extraction capabilities compared to holes. This finding is surprising because electrons, which are minor carriers, are heavily influenced by empty traps in *p*-type doped $FASnI_3$. Therefore, one would expect a lower amplitude of trSPV signals for ETLs interfaces than HTLs interfaces. This observation directly highlights one of the bottlenecks in the state-of-the-art TPSC-HTL interface.

To investigate electron and hole extraction simultaneously, we compared the trSPV signals for perovskite samples with ETLs and HTLs, as depicted in **Figure SI 13**. The results demonstrate a significant enhancement in the extracted charges, with the trSPV signal's amplitude (-160 mV) several times larger than the combined signals of the pure ETL and HTL. This notable increase in the trSPV peak amplitude signifies the strong correlation between electron and hole extraction and highlights the substantial recombination losses caused by unextracted carriers. As carrier recombination occurs through various channels, including non-radiative, surface, and radiative pathways, **supplementary note 1 and Figure SI 14** provide a more in-depth discussion.

### Simulation of charge extraction and recombination in tin perovskite interfaces:

In previous chapters, we qualitatively demonstrated charge transport dynamics in tin perovskite interfaces using various ETLs and HTLs. In this section, we provide a quantitative analysis of charge extraction efficiency and the mechanisms responsible for charge loss by conducting charge extraction and recombination simulations. We simulate charge transport at interfaces that exhibit the highest charge extraction efficiency in tin perovskites and compare the quality of charge extraction with similar interfaces of lead perovskites, $(Cs_{0.05}FA_{0.855}MA_{0.095})Pb(I_{0.9}Br_{0.1})_3$ with a power conversion efficiency of 22%. This comparison gives insights into the primary mechanisms causing charge losses and provides valuable information for developing future strategies to enhance tin perovskites and other optoelectronic devices.

There are two channels for charge loss: radiative and non-radiative recombination. The radiative recombination rate ($C_b \cdot \Delta n(t) \cdot (\Delta p(t) + p0)$) is influenced by the background hole concentration $p0$. The total non-radiative recombination ($\frac{\Delta n}{\tau_{non-rad_e}}$ and $\frac{\Delta p}{\tau_{non-rad_h}}$) consists of bulk ($\frac{\Delta n}{\tau_{bulk_e}}$), surface (($s_e \cdot \Delta$n and $s_h \cdot \Delta$p), and interface ($s_{ie} \cdot \Delta$n and $s_{ih} \cdot \Delta$p) contributions, as demonstrated in **Equation S1-2**, where $\Delta$n(t) and $\Delta$p(t) represent photogenerated electrons and holes, respectively. $s_e$ and $s_h$ are non-radiative surface recombination rates at the perovskite surface and $s_{ie}$ and $s_{ih}$ at the HTL and ETL interface. We consider these recombination channels as key factors governing charge dynamics in perovskite interfaces, along with electron and hole extraction rates $K$e·$n$(t) and $K$h·$p$(t), where $K$e and $K$h denote the coefficients for electron and hole extraction rates, respectively, as shown in **Equations 1-2** and **Figure 4a**. Further details regarding the model can be found in SI equations **S1-S4**.

$$\frac{d\Delta p}{dt} = -K_h \Delta p + K_{hb} \Delta p_{HTM} - C_b\big((\Delta p + p0)\Delta n\big) - \frac{\Delta p}{\tau_{non-rad_h}} \quad \textbf{(1)}$$

$$\frac{d\Delta n}{dt} = -K_e \Delta n + K_{eb} \Delta n_{HTM} - C_b\big((\Delta p + p0)\Delta n\big) - \frac{\Delta n}{\tau_{non-rad_e}} \quad \textbf{(2)}$$

We initially employed the Levenberg-Marquardt method to fit the experimental data of $FASnI_3/C_{60}$ and PEDOT:PSS/$FASnI_3$, aiming to determine the rate constants that best align with the experimental trSPV results.[34, 40] The fit results are presented in **Figure 4b** and **Table S1**, while the time evolution of electrons, holes, and extracted electrons is depicted in **Figure 4c**. Subsequently, we performed a similar fitting process for the electron and hole extraction dynamics in lead perovskites. We utilized trSPV measurements conducted under identical conditions, as shown in **Figures 4e and f** and **Table S1**. Careful examination of the charge extraction dynamics reveals a direct correlation between the rise of the trSPV signal in time (**Figures 4b and e**) and a reduction in carrier concentration (**Figures 4c and f**) within the perovskite film. This reduction is a consequence of the charge extraction process, increasing the extracted electron concentration in ETL and HTL. The further decay of trSPV and the maximum amplitude of the trSPV signals are significantly influenced by radiative and non-radiative recombination processes that tend to diminish the overall carrier density.

The simulation results show significant charge extraction and recombination differences between lead and tin perovskites. The lead perovskite system exhibits lower recombination losses, resulting in a longer carrier lifetime, higher concentration of extracted charges, and an extended exponential tail, as indicated in **Figure 4f**. Moreover, the lead perovskite demonstrates a larger total extracted charge, $Q_{ex}$, calculated as the surface area under $n_{ex}(t)$ and $p_{ex}(t)$, as summarized in **Figure 4d**. Interestingly, the total extracted charge in the lead perovskite system is nearly balanced, with $Q_e/Q_h = 2$. However, a much more significant difference is observed in the tin perovskite system, with $Q_e/Q_h = 14.5$. Additionally, the tin perovskite system loses a considerable number of extracted charges. Specifically, the ETL extracts 30 times fewer electrons, and the HTL extracts 160 fewer holes (at the exact condition of 5 ns laser pulse excitation) than the lead perovskite system. This significant reduction in extracted charge highlights the poor capabilities of tin perovskite solar cell interfaces in extracting both holes and electrons and significant recombination losses.

After demonstrating the humble charge extraction capabilities of tin perovskites compared to lead perovskites interfaces, next, we quantify each recombination channel's impact on the extracted charges. Three factors can influence the extracted charges: the extraction rate constant (rate constant deficit), non-radiative recombination, and radiative recombination. We considered the charge extracted from lead perovskite as the baseline. We examined the impact of factors affecting charge extraction individually to assess their influence on charge losses, which would reveal a road map to tackle the energy bands misalignment challenge in tin perovskites. We found that tin perovskites exhibit lower charge extraction constants, as illustrated in **Figure 4g**. The rate constraints for the HTL and ETL are 33 and 12 times lower than those in Pb perovskite interfaces, respectively. This difference may arise from the energetic misalignment between the absorber layer and the CTLs, as highlighted in **Figure 1**. Alternatively, it could be attributed to the damage on the perovskite surface, resulting in a poorly aligned interface with the CTLs. The charge losses are depicted in **Figures 4h and i**, where 100% represents the electron and hole charge from **Figure 4d**, respectively. The charge loss percentage is calculated using the formula loss = $(Q_{lead} - Q_{tin}) / Q_{lead}$.

In both ETLs and HTLs, the primary source of charge loss is non-radiative recombination, accounting for 90% of the charge loss in the ETL interface and 99% in the HTL interface. This slightly higher loss percentage at the HTL interface can be attributed to the larger surface recombination, as indicated in **Tables S1-2**. The extraction rate constant deficit contributes to 70 and 60% of charge loss in the ETL and HTL interfaces, respectively. Interestingly, at the HTL interface, the high material doping ($1.5\times 10^{17}$ $cm^{-3}$) results in significant hole losses, with 97% of holes being lost. In contrast, the same doping level in the ETL induces only 28% of electron loss at its interface. To identify the reasons for such behaviour, we plotted quasi-Fermi level splitting (QFLS) in both lead and tin systems, demonstrating the effect of recombination channels on charge concentration in the perovskite absorber film itself, considering additional ETL and HTL surface recombination (**Figure 4j**). Due to the significantly enhanced passivation capability of 2PACz in lead perovskites, QFLS reaches a value of 1 eV (after recalculation of QFLS to the bandgap of $FASnI_3$), which is very close to the Shockley-Queisser limit of 1.1 eV in $FASnI_3$. In contrast, the $C_{60}$ ETL interface is more susceptible to non-radiative recombination, resulting in a QFLS of 0.84 eV. Consequently, doping and radiative recombination substantially influence the charge extraction potential in the HTL interface in tin perovskites compared to the ETL interface.

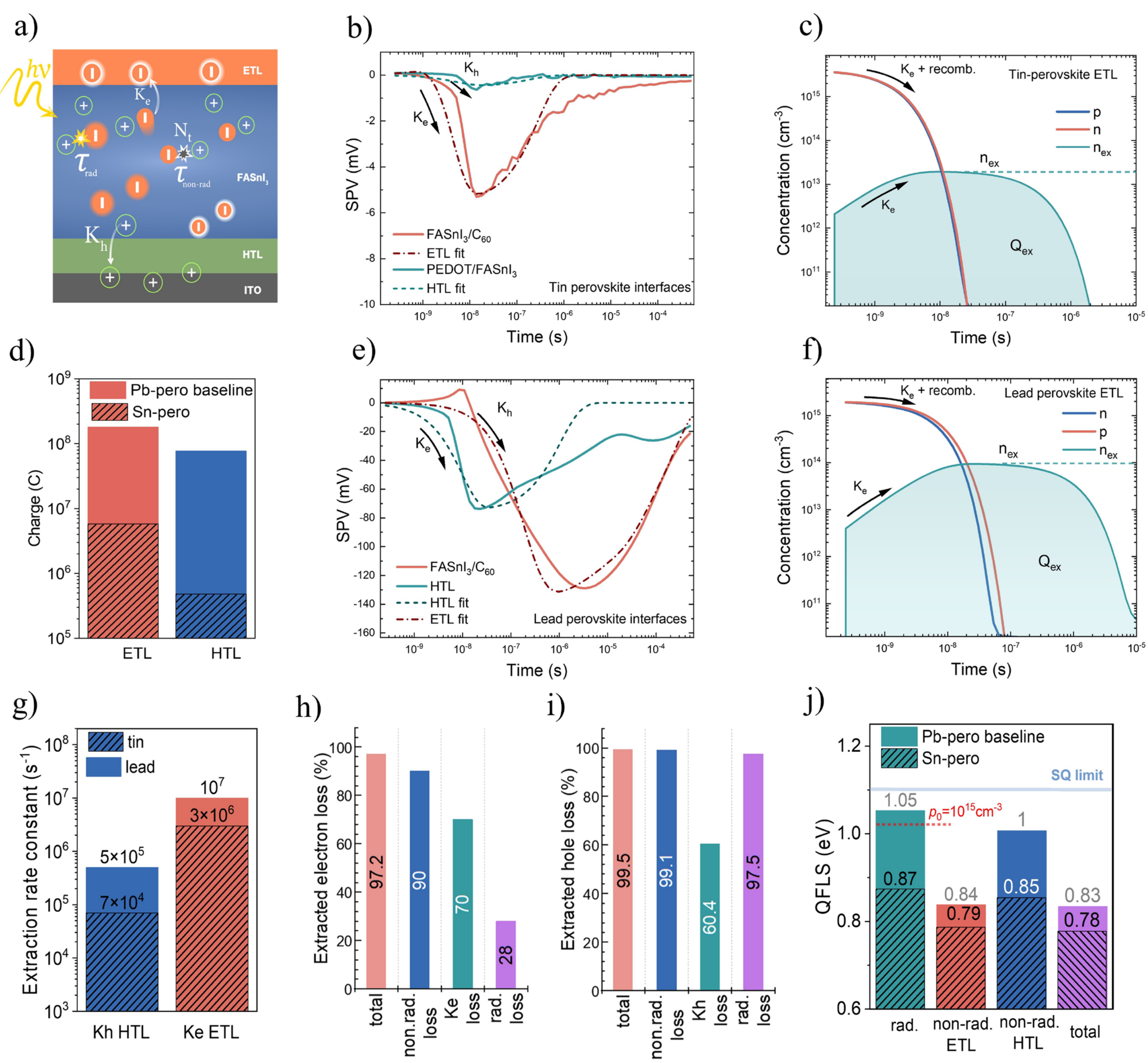

**Figure 4: Non-radiative recombination is the primary charge loss channel in TPSCs according to the simulation of charge extraction and charge losses in $FASnI_3$ and $(Cs_{0.05}FA_{0.855}MA_{0.095})Pb(I_{0.9}Br_{0.1})_3$ perovskite systems.** (a) A theoretical model describing charge extraction. (b) Fitting of experimental data in tin perovskites ETLs and HTLs interfaces. (c) Fitting of experimental data in lead perovskite ETL and HTL interfaces. (d) Total electron and hole charge extracted by HTL and ETL per one laser pulse. (e) and (f) Time evolution of electron and hole concentrations in the perovskite layer, along with the evolution of extracted electrons in the tin and lead systems, respectively. (g) Hole and electron extraction rate constants. (h) and (i) Loss of extracted electrons and holes induced by non-radiative recombination, extraction constant deficit, and

radiative recombination in the ETL and HTL, respectively, using the charge in the Pb system as a baseline. (j) Quasi-Fermi Level Splitting (QFLS) affected by non-radiative and radiative recombination channels.

Several key steps must be taken to achieve the baseline performance of lead perovskite (22% efficiency). First, it is crucial to significantly reduce non-radiative recombination and passivate trap states both in bulk and on the surface of the absorber tin perovskite material. This is necessary to achieve lifetimes in the microsecond range, similar to those observed in lead perovskites. Secondly, the background hole density needs to be reduced to at least the level of $10^{15}$ $cm^{-3}$. Thirdly, there is a need to improve the alignment of the ETL and HTL to enhance the extraction rate constants, aiming for values around $10^7$ $cm^{-3}$ $s^{-1}$.

## Simulation of tin perovskite solar cell operation:

We have comprehensively studied the factors influencing energy alignment, charge extraction, and recombination in tin perovskite interfaces. We further investigated the impact of these limiting factors on a representative TPSC device with a PCE of 6%. We used 2D drift-diffusion simulations under steady-state illumination conditions to model the device. Based on the findings of KP and PYS measurements, we modelled the perovskite, $C_{60}$, and PEDOT:PSS layers as crystalline semiconductors, considering their respective thicknesses, energy bandgaps, and electron affinities. The perovskite *p*-type doping was set to $1.5 \times \times 10^{17}$ $cm^{-3}$, as determined in the present study by the Hall, PL, and KP-PYS measurements. Additionally, we considered the significant valence band offset (VBO) of approximately 0.47 eV between the perovskite layer and $C_{60}$, as revealed by the KP-PYS study.

To facilitate the injection of photogenerated holes from the perovskite to PEDOT:PSS via thermionic emission over the barrier, we adjusted the VBM of PEDOT:PSS in our model. We increased the VBM by 0.12 eV compared to our experimental data, thereby modifying $E_{i\text{-PEDOT}}$ (from $E_{i\text{-exp}}$=5.30 eV to $E_{i\text{-model}}$=5.18 eV), considering the possibility of defect-assisted tunneling across the barrier. Furthermore, we note that the VBM edge of PEDOT:PSS may shift upon perovskite deposition. Ohmic contacts were implemented at the $C_{60}$ and PEDOT:PSS sides of the solar cell, with corresponding WF values of 3.95 and 4.74 eV, measured for Ag:BCP and ITO, respectively.

Given the considerable CBO of approximately -0.53 eV between the perovskite and $C_{60}$, we anticipated significant non-radiative recombination at this interface. To address this issue, we introduced a fictitious layer between the two materials with Shockley-Read-Hall non-radiative recombination. This 1 nm-thick layer shared the CBM of $C_{60}$ and the VBM of the perovskite, enabling recombination between electrons in $C_{60}$ and holes in the perovskite. This additional layer was necessary to align the $V_{OC}$ of our simulated perovskite solar cell with the experimental data. We plotted the *J-V* curve under 1-sun illumination of our simulated device (in blue) and the experimental points in **Figure 5a**. We report all the material parameters employed to produce the blue *J-V* curve in **Table S5**.

**Figure 5b** illustrates the equilibrium band diagram of the material stack. As mentioned earlier, a significant cliff at the $FASnI_3/C_{60}$ interface limits the maximum achievable $V_{OC}$. Since the perovskite doping density exceeds that of $C_{60}$, most of the built-in voltage drops across the $C_{60}$ film, resulting in an ineffective separation of photogenerated carriers due to a negligible electric field strength across the perovskite layer. Notably, with a field-free region, most carriers are generated near the PEDOT:PSS/$FASnI_3$ interface. Therefore, electron collection and extraction at the back contact is hindered, considering the low lifetime of the free carriers in $FASnI_3$.

**Figure 5a** also contains the *J-V* curve obtained by illuminating the material stack from the $C_{60}$ side (red curve) to test the feasibility of the *n-p-p* solar cell architecture. In contrast, all the material parameters are equal to the *n-p-p* configuration. If the solar cell was illuminated from the ETL side, most of the electrons would be photogenerated near the $C_{60}$ interface, reducing the impact of bulk non-radiative recombination on the electrons and constituting the minority carriers in our perovskite layer. Therefore, we found that the n-p-p architecture delivers a larger PCE value than the p-p-n structure ($PCE_{n-p-p}$ = 7.2% compared to $PCE_{n-p-p}$ = 6.2%).

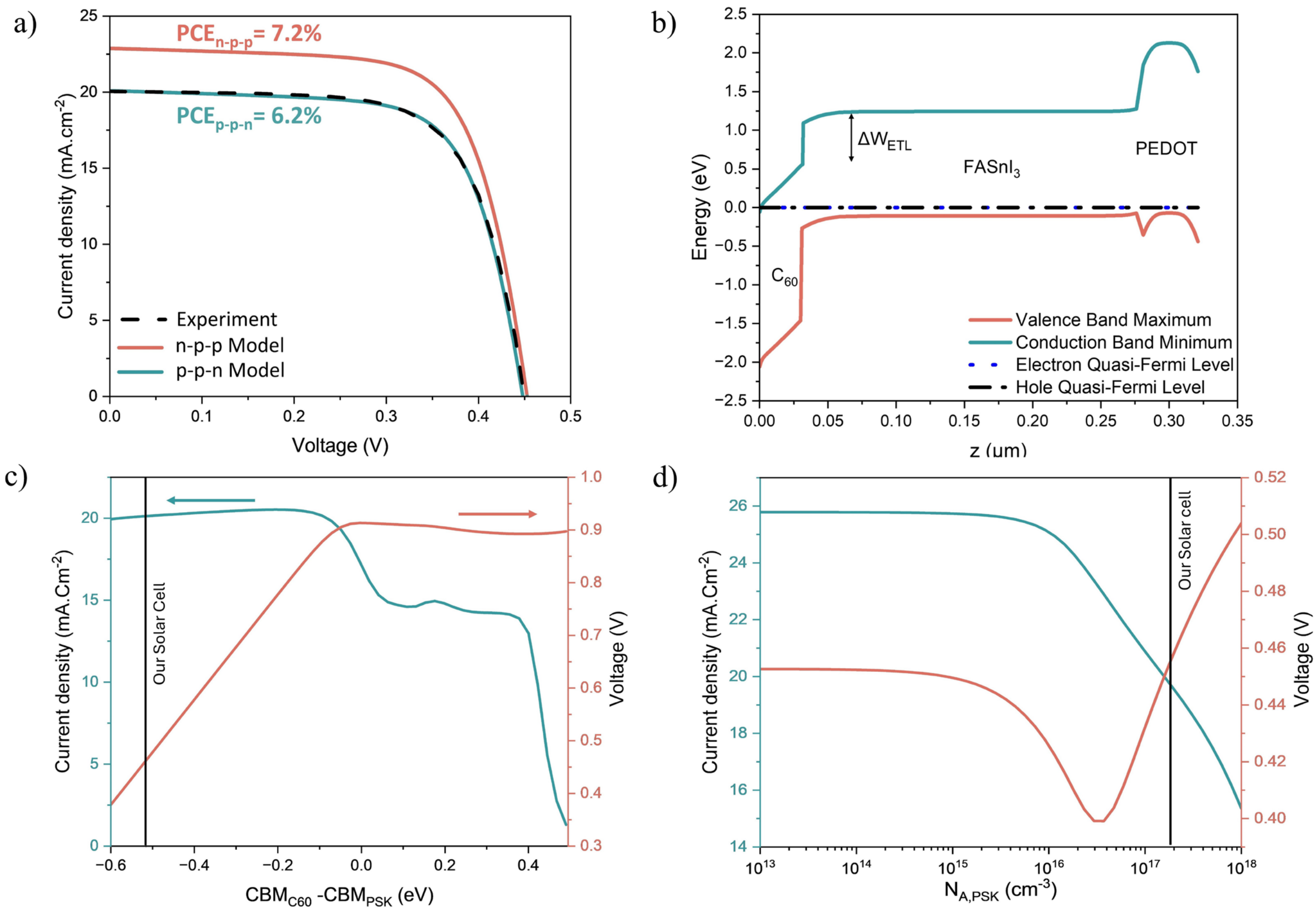

**Figure 5**: **CBO, non-radiative recombination, and doping limits $FASnI_3$ solar cell performance in p-p-n and n-p-p architecture** (a) *J-V* curves of the simulated solar cell in *p-p-n* and *n-p-p* configuration with respect to the experimental p-i-n device. (b) The equilibrium band diagram of the solar cell employs the parameters

listed in **Table S5**. (c) $J_{SC}$ and $V_{OC}$ of the solar cell as a function of the conduction band offset between the perovskite and $C_{60}$, and (d) as a function of the perovskite *p*-type doping.

After achieving a satisfactory fit for the *J-V* curve, we explored variations of energy offset, doping, and recombination rate on the TPSCs efficiency, demonstrating room for further improvement. First, we focused on the CBO at the $FASnI_3/C_{60}$ interface. **Figure 5c** shows how TPSCs would perform if we mitigated the large CBO. Starting from the conditions indicated by the solid black vertical line ($\Delta W_{ETL} = CBM_{C60} - CBM_{psk} = -0.53$ eV), $J_{SC}$ would not degrade significantly, while $V_{OC}$ would nearly double up until $CBM_{C60} \simeq CBM_{psk} - 0.1$ eV. Increasing the energy gap between $CBM_{C60}$ and $VBM_{psk}$ would reduce non-radiative recombination between electrons in $C_{60}$ and holes in the perovskite layer, benefiting the $V_{OC}$. Importantly, this improvement can be achieved without decreasing the non-radiative lifetime of the interface layer (or its trap density), which we maintained at 1.5 ns in our simulations. If the CBO increases further, a barrier for electron extraction emerges, severely degrading $J_{SC}$. On the contrary, $V_{OC}$ would not be affected, as it is determined under zero-current conditions.

Next, we consider the effect of *p*-doping density on TPSC $J_{SC}$ and $V_{OC}$. Starting from the experimental concentration indicated by the solid black vertical line ($N_{A,\ pero} = 1.5 \times 10^{17}$ $cm^{-3}$), reducing the perovskite doping to values below $10^{16}$ $cm^{-3}$ establishes a constant electric field[41] throughout the perovskite film. This field improves charge extraction, enhances $J_{SC}$, and minimizes bulk non-radiative recombination. However, $V_{OC}$ is slightly affected since it is primarily limited by interface recombination. The non-monotonous trend of $V_{OC}$ around 0.4 V can be attributed to two competing phenomena. On the one hand, increased perovskite doping negatively impacts $V_{OC}$ since it further reduces the electric field strength across the perovskite layer. On the other hand, higher perovskite doping values may slightly benefit $V_{OC}$ since non-radiative recombination at the $FASnI_3/C_{60}$ interface would decrease. As the perovskite doping density increases, the interface Fermi level shifts from values close to the $C_{60}$ CBM toward the perovskite VBM, as illustrated in **Figure SI 15**. When the interface Fermi level approaches midgap, the non-radiative recombination rate at the interface increases since the populations of holes and electrons become comparable.

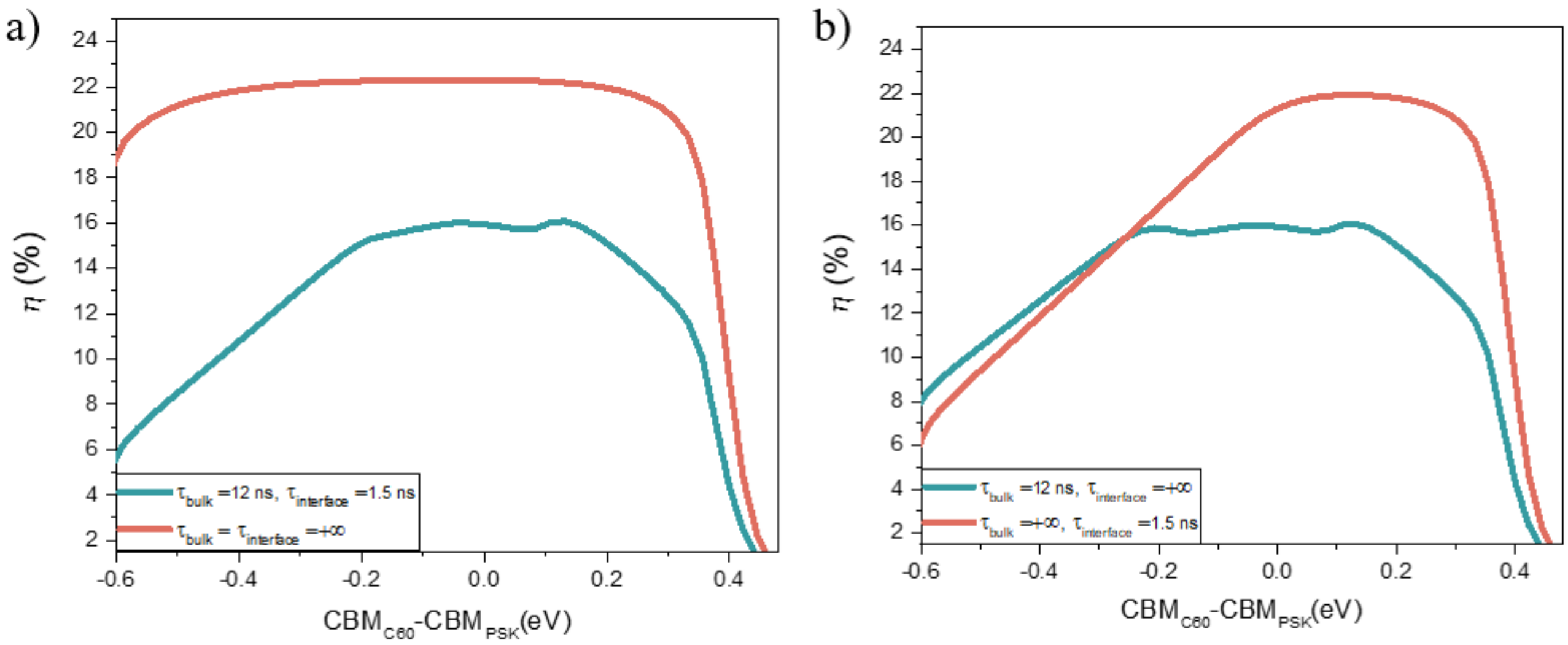

**Figure 6:** PCE of $FASnI_3$ devices as a function of $C_{60}$/perovskite layer CBO (in eV) for a p-type doping concentration of $1\times10^{15}$ $cm^{-3}$.

To visualize the impact of non-radiative recombination on tin perovskite solar cells, we calculated the PCE with and without the non-radiative recombination rate determined for our devices, as illustrated in **Figure 6a**. The passivation of recombination channels results in a 22% improvement in PCE, highlighting the potential of TPSCs. Without addressing non-radiative recombination, but only by resolving the ETL, we can achieve a PCE of 16%. Furthermore, we calculated the PCE by activating surface and bulk recombination individually to elucidate the contributions of surface and bulk recombination, as depicted in **Figure 6b**. Our analysis reveals that bulk recombination contributes dominantly to non-radiative recombination in tin perovskite solar cells. If we resolve the interface offset, the main limitation factor of TSPCs is bulk non-rad recombination. The doping concentration of $1\times 10^{17}$ itself leads only to a minor change in PCE as shown in **Figure SI 16**. Therefore, passivation and engineering of bulk defects must be considered as the primary step to achieve a PCE exceeding 20% in tin perovskite photovoltaic systems.

## Conclusion:

In this work, we studied the electrical, transport, and defect properties of the $FASnI_3$-based tin devices. We investigated the efficiency limiting factors of the perovskite $FASnI_3$-based solar cells. We revealed the development of their photovoltaic potential in a layer-by-layer completion of the ITO/PEDOT/$FASnI_3$/$C_{60}$/BCP/Ag device stack. We found that the doping concentration of a standard $FASnI_3$ device using DMSO as a solvent is $1.5\times10^{17}$ $cm^{-3}$. The energy band diagram of tin perovskite devices reveals an energy offset of $\Delta E_v = 0.42$ eV at the HTL interface and another negative band offset of $\Delta E_c = -0.30$ eV at the ETL interface. Those offsets lead to high interfacial recombination and low charge extraction, reducing device efficiency. Consequently, better-aligned CTLs are needed for more efficient tin perovskites.

The careful layer-by-layer investigation of the device stack revealed that BCP dramatically enhances electron extraction and contributes considerably to the $V_{OC}$ of the device under investigation. The early molecules of BCP blend with silver as a metal contact to form a charge transfer complex with an intermediate energy level of 4 eV, which is in an optimum position for electron extraction from the $C_{60}$ to the metal contact.

Charge extraction investigations using trSPV revealed a low extraction rate in the HTL interface with a high recombination rate close to the interface compared to the bulk. In addition, the charge extraction at the $FASnI_3$/PEDOT:PSS interface is driven mainly by two processes: hole extraction and electron trapping due to energy band misalignment. On the other hand, electrons show better extraction dynamics than holes, with superior performance for $C_{60}$ over ICBA and PCBM due to the higher electron mobility regardless of the higher CBO.

Comparing tin and lead perovskites charge extraction and recombination differences shows that lead has 30 times higher electron extraction and 160 times higher hole extraction than tin perovskite. This is mainly due to the lower extraction constants due to energetic misalignment, which indicates the importance of enhancing the energy alignment at both interfaces. Finally, the current *p-p-n* structure of the tin devices leads to high interfacial charge losses because most of the charge excitations occur close to the HTL interface that has a very high band offset; however, shifting to the n-p-p structure would be more efficient due to the better extraction at the ETL interface.

We suggest designing a new charge extraction system well aligned with the tin perovskite film or trying the opposite strategy of controlling the WF of tin perovskites by polar surface additives[42] to enhance the band alignment with good CTLs such as $C_{60}$/BCP. The methodology and findings presented in this study can be applied to identify optimal ETL and HTL candidates that enable tin perovskites to exceed the efficiency threshold of 20% and increase the commercialization possibilities of such green solar cell material.